\documentstyle[prc,aps,preprint,psfig,amsmath]{revtex}

\begin{document}
\title{Photoproduction of mesons in nuclei \\ at GeV energies
\thanks{Work supported by DFG and BMBF}}
\author{M. Effenberger and U. Mosel \\[5mm]
Institut f\"ur Theoretische Physik,
Universit\"at Giessen\\D-35392 Giessen, Germany \\
UGI-99-21}
\maketitle

\tightenlines

\begin{abstract}
In a transport model
that combines initial state interactions of the photon with final state
interactions of the produced particles
we present a calculation
of inclusive photoproduction of mesons in nuclei in the energy range from
1 to 7 GeV. We give predictions for the photoproduction cross sections of
pions, etas, kaons, antikaons,
and $\pi^+\pi^-$ invariant mass spectra in $^{12}$C and $^{208}$Pb. The effects
of nuclear shadowing and final state interaction of the produced particles
are discussed in detail.
\end{abstract}

\vspace*{1cm}
\noindent

\noindent

\newpage
%--------------------------------------------------------------------------
\section{Introduction}
Photoproduction of mesons in nuclei offers the possibility to study the
interaction of photons with nucleons in the nuclear medium. In the total
photonuclear absorption cross section one observes experimentally
for photon energies above
1 GeV a reduction of the absorption strength in nuclei which is known as
shadowing effect. While Vector Meson Dominance (VMD) models are able to
quantitatively describe this effect for photon energies above about 4 GeV
they usually underestimate the effect for lower photon energies (see e.g.
Ref.~\cite{Muccifora98}). In Ref.~\cite{Bianchi99} it has been speculated
that this may be related to a decrease of the $\rho$-meson mass in the
nuclear medium.
\par The in-medium properties of the $\rho$-meson have found
widespread interest during the past decade as they may be related to
chiral symmetry \cite{Mosel91,Ko97}.
The experimental data on dilepton production in
nucleus-nucleus collisions at SPS energies \cite{Agakichiev95,Mazzoni94}
also seem to indicate a change of the $\rho$-meson
spectral function (lowering of the mass or broadening) in the nuclear
medium \cite{Li96,Cassing99}.
\par A third experimental observation that may be
related to in-medium properties of the $\rho$-meson is the disappearance
of the $D_{13}(1520)$-resonance in the total photoabsorption cross section
in nuclei at photon energies around 800 MeV. In Refs.~\cite{abspaper,Peters98}
it has been proposed that this disappearance is caused by the large coupling
of the $D_{13}$ to the $N \rho$-channel and a medium modification of the
spectral function of the $\rho$-meson.
\par While a consistent theoretical descriptions of these observations is not
yet available our transport approach is a first step into this direction.
During the past years we have developed a semi-classical BUU transport model
that allows us to calculate inclusive particle production in heavy-ion
collisions from 200 AMeV to 200 AGeV, in photon, and in pion induced reactions
with the very same physical input.
This model has already been very successfully
applied to the description of heavy-ion collisions at SIS energies
\cite{Teis97,Hombach99} and photoproduction of pions and etas up to 800 MeV
\cite{prodpaper}. Just recently we have given predictions for dilepton
production in pion-nucleus reactions \cite{pidil}, that will be measured
by the HADES collaboration at GSI \cite{HADES}, and in photon-nucleus reaction
in the energy range from 800 MeV to 2.2 GeV \cite{dilgam} that is accessible
at
TJNAF \cite{TJNAF}. In these studies we have investigated direct observable
consequences of medium modifications of the vector mesons through their
dilepton decay.
\par The calculation of photoproduction in the region of the nucleon resonances
(up to photon energies of about 1 GeV) has recently been extended to
electroproduction \cite{lehr}. In this study we have also discussed different
scenarios that might lead to a disappearance of the $D_{13}$ in the total
photoabsorption cross section with respect to their influence on more
exclusive observables.
\par It is the purpose of the present paper to give predictions for pion,
eta, kaon, and antikaon production in photon-nucleus reactions in the energy
range from 1 to 7 GeV. A comparison of these calculations with experiments
that are possible at ELSA and TJNAF will on the one hand help to
improve our understanding of the onset of the shadowing effect. On the other
hand is photoproduction an excellent tool to study medium modifications of
the produced mesons since the final observables get strongly modified by
the final state interaction of the primary produced particles with the
nuclear medium. In photon-nucleus reactions the
produced particles have in general larger momenta with respect to the
nuclear environment than in heavy-ion collisions. Therefore, photonuclear
experiments yield partly complementary information about the in-medium
self energies of mesons compared to heavy-ion experiments.
They also have the great advantage that they allow to study the properties
of the produced hadrons in an environment that is much closer to
equilibrium than in a heavy-ion collision.
\par While the early seventies have already seen a remarkable number of
experimental and theoretical studies of photoinduced particle production
(for a rather complete review see Ref.~\cite{bauer}) the emphasis there
was on the study of coherent production processes. Incoherent production
had to rely for an interpretation on the Glauber approximation that
involves a number of approximations and restrictions. The study described
here is free of most of these and thus provides a more reliable framework
for an interpretations of such reactions. Unfortunately,
the few existing experimental data
in Refs.~\cite{boyarski69,meyer72,mcclellan69} on incoherent photoproduction
of mesons in nuclei
were all obtained under very restrictive experimental conditions which,
furthermore, can not be reconstructed from the literature. We can, therefore,
not compare our results to any data.
\par Our paper is organized as follows: In Section~\ref{sec:buu} we describe
briefly the BUU transport model.
In Section~\ref{sec:ele} we discuss our treatment of elementary
photon-nucleon collisions and, in particular, the implementation of the
shadowing effect.
Our results for photoproduction of pions, etas,
kaons, and antikaons are presented in Section~\ref{sec:results}.
We close with a summary in Section~\ref{sec:sum}.

\section{The BUU model}
\label{sec:buu}

We use a semi-classical transport model (BUU) for a description of the
final state interactions (fsi) of the produced particles. This description
allows a full coupled channel treatment of the fsi, including non-forward
processes.
Our model has been described in Ref.~\cite{dilgam}. Therefore we restrict here
ourselves to a brief presentation of the basic ideas and only discuss in
detail the basic new features of our method.
\par The BUU equation describes the classical time evolution of a
many-particle system under the influence of a self-consistent mean field
potential and a collision term. For the case of identical particles it is given
by:
\begin{equation}
(\frac{\partial}{\partial t} + \frac{\partial H}{\partial \vec{p}}
\frac{\partial}{\partial \vec{r}} - \frac{\partial H}{\partial \vec{r}}
\frac{\partial}{\partial \vec{p}})f=I_{coll}[f] \quad,
\label{transport}
\end{equation}
where $f(\vec{r},\vec{p},t)$ denotes the one-particle phase space density
with $\vec{r}$ and $\vec{p}$ being the spatial and momentum coordinates of the
particle. $I_{coll}$ is the collision term and $H(\vec{r},\vec{p},f)$
stands for the single particle mean field
Hamilton function which, in our numerical realization~\cite{Teis97},
is given as:
\begin{equation}
H=\sqrt{(\mu+S)^2+\vec{p}^2} \quad,
\end{equation}
where $S(\vec{r}, \vec{p}, f)$ is an effective scalar potential.
For a system of non-identical particles one gets an equation for each particle
species that is coupled to all others by the collision term and/or the mean
field potential. Besides the nucleon we
take into account all nucleonic resonances that are rated with at least
2 stars in
Ref.~\cite{Manley}: $P_{33}$(1232), $P_{11}$(1440), $D_{13}$(1520), $S_{11}$(1535), $P_{33}$(1600), $S_{31}$(1620), $S_{11}$(1650), $D_{15}$(1675), $F_{15}$(1680), $P_{13}$(1879), $S_{31}$(1900), $F_{35}$(1905), $P_{31}$(1910), $D_{35}$(1930), $F_{37}$(19

50), $F_{17}$(1990), $G_{17}$(2190), $D_{35}$(2350). The resonances couple to the following channels: $N \pi$, $N \eta$, $N \omega$, $\Lambda K$, $\Delta(1232) \pi$, $N \rho$, $N \sigma$, $N(1440) \pi$, $\Delta(1232) \rho$.
\par We also propagate explicitly the following baryonic resonances with total
strangeness $S=-1$:
$\Lambda$, $\Sigma$, $\Sigma(1385)$, $\Lambda(1405)$, $\Lambda(1520)$,
$\Lambda(1600)$, $\Sigma(1660)$, $\Lambda(1670)$,
$\Sigma(1670)$, $\Lambda(1690)$,
$\Sigma(1750)$, $\Sigma(1775)$, $\Lambda(1800)$,
$\Lambda(1810)$, $\Lambda(1820)$, $\Lambda(1830)$,
$\Lambda(1890)$, $\Sigma(1915)$, $\Sigma(2030)$, $\Lambda(2100)$,
$\Lambda(2110)$. The parameters of these resonances are consistent with
the values given by the PDG~\cite{PDG} and are listed in
Table~\ref{table1}. The resonances  couple to the channels:
$\Lambda \pi$, $\Sigma \pi$,
$\Sigma(1385) \pi$, $\Lambda \eta$, $N \bar{K}^*(892)$, $\Lambda(1520) \pi$.
The mass dependence of the partial decay widths is treated analogous to the
case of the nucleonic resonances (see Ref.~\cite{dilgam}); in cases when
the relative angular momentum of the decay products is not uniquely given
by the quantum numbers we use the lowest possible one. In the mesonic
sector we take into account the following particles: $\pi$, $\eta$, $\rho$,
$\omega$, $\sigma$, $\phi$, $K$, $\bar{K}$, $K^*(892)$, $\bar{K}^*(892)$.
\par For a detailed description of the used cross sections in the
non-strange sector we refer to Refs.~\cite{dilgam,Teis97}.
Baryon-baryon collisions above invariant energies of 2.6 GeV and meson-baryon
collisions
above 2.2 GeV are described by using the string fragmentation model
FRITIOF~\cite{FRITIOF}.
For strangeness
production in low energy pion-nucleon collisions we adopt the parameterizations
for $\pi N \to \Lambda K$ and $\pi N \to \Sigma K$ from Ref.~\cite{Huang}
and for $\pi N \to N K \bar{K}$ from Ref.~\cite{Sibirtsev}. Strangeness
production in baryon-baryon collisions is only of minor importance for the
calculations presented here and is described in Ref.~\cite{schaffner}.
The cross sections for the interactions of kaons and antikaons with nucleons
can be found in Appendix~\ref{app:kaon}.

\section{Photon-nucleon interaction}
\label{sec:ele}
\begin{figure}[h]
\centerline{\psfig{figure=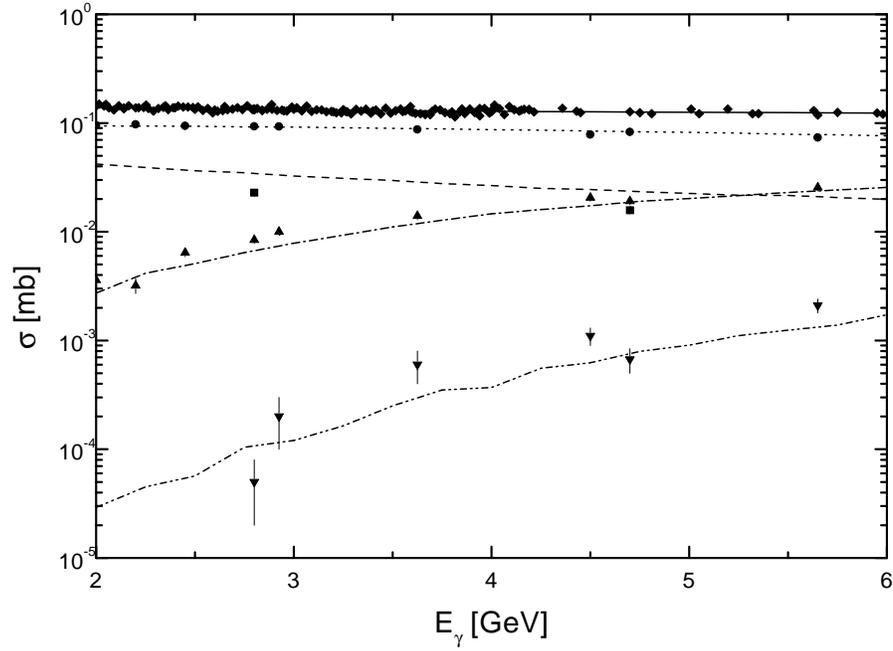,width=15cm}}
\caption{Charged particle multiplicity cross sections in $\gamma p$ reactions:
$\gamma p \to$ 1 charged particle (dashed line (calculation as described in the text), 
squares (experimental data from Ref.~\protect\cite{landolt})), 3 charged particles (dotted line, circles), 5 charged particles (dot-dashed line, up triangles), 
7 charged particles (dot-dot-dashed line, down triangles). Also shown is the total cross section (solid line, rhombs).}
\label{chargemult}
\end{figure}

In Refs.~\cite{dilgam,prodpaper} we have described in detail how we
calculate photonuclear reactions. The total cross section for any observable
is given as an incoherent sum over the contributions from all
nucleons in the nucleus where the final state interactions of the particles
produced in the primary $\gamma N$ collisions are calculated using the
transport equation~(\ref{transport}). For invariant energies below 2.1 GeV
(corresponding to $E_\gamma=1.88$ on a free nucleon at rest) we describe
elementary $\gamma N$ collisions as in Ref.~\cite{prodpaper} by an explicit
calculation of the cross sections for production of nucleonic resonances as
well as one-pion, two-pion, eta, vector meson, and strangeness
production. For larger
energies we use the string fragmentation model FRITIOF~\cite{FRITIOF} where
we initialize a zero mass $\rho^0$-meson for the photon following a VMD
picture. In Fig.~\ref{chargemult} we show that this procedure gives an
excellent description of charged particle multiplicities in photon-proton
collisions. The agreement seen there is better than could be expected from
a model that had been developed for applications at higher energies. However,
we do not expect the Lund model to give correct predictions for all specific
channels, especially with respect to isospin. Since in the Lund model
flavor exchange mechanisms are not included processes like, e.g.
$\rho N \to \Lambda K$, are not possible. Therefore we treat exclusive
strangeness production independent of the Lund model also for energies
above the string threshold. The used cross sections for exclusive
strangeness production are discussed in Appendix~\ref{app:str}.
\par Nuclear shadowing of the incoming photon is taken into account for photon
energies above 1 GeV by adopting the model of Ref.~\cite{bauer} in the
following way. For the total photon-nucleus cross section we have:
\begin{equation}
\label{totshad}
\sigma_{\gamma A}=A \sigma_{\gamma N}-\int d^3r \rho(\vec{r}) S(\vec{r})
\equiv A_{eff} \sigma_{\gamma N} \quad,
\end{equation}
where $\rho(\vec{r})$ denotes the nuclear density and $S(\vec{r})$ is
given as:
\begin{equation}
S(b,z)= -8 \pi  \sum_V \left( T_{\gamma V}\right)^2
{\rm Im} \left\{ i \int_{-\infty}^z dz^\prime \rho(b,z^\prime)
\exp \left[i q_\parallel^V(z^\prime-z)+2 i \sqrt{\pi} T_{V V}
\int_{z^\prime}^z \rho(b,\xi) d\xi \right] \right\}\quad,
\label{sbz}
\end{equation}
where we have expressed everything in cylindrical coordinates $(b,z)$ with
the photon momentum along the $z$-axis. The momentum transfer $q_\parallel^V$
is:
\begin{equation}
q_\parallel^V=E_\gamma-\sqrt{E_\gamma^2-M_V^2} \quad,
\end{equation}
with $E_\gamma$ and $M_V$ denoting the photon energy and the mass of the
vector meson V, respectively. In Eq.~(\ref{sbz}) $T_{a b}$ is the amplitude
for the process $a N \to b N$. We used all parameters from \cite{bauer}
(model I) taking into account $\rho$, $\omega$ and $\phi$ mesons.
\par
\begin{figure}[h]
\centerline{\psfig{figure=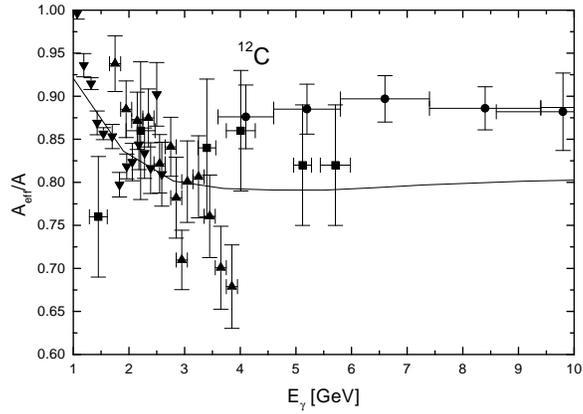,width=9.5cm}}
\caption{Effective mass number as function of photon energy for
photoabsorption on $^{12}$C. The experimental data are taken from Refs.
\protect\cite{caldwell} (circles), \protect\cite{heynen} (squares), \protect\cite{brookes} (up triangles), \protect\cite{Muccifora98} (down triangles). 
The latter two are obtained from the
experimental data on the total cross sections by using the parameterization of the cross section on a single nucleon from Ref.~\protect\cite{PDG94}.}
\label{shad_12c}
\end{figure}

\begin{figure}[h]
\centerline{\psfig{figure=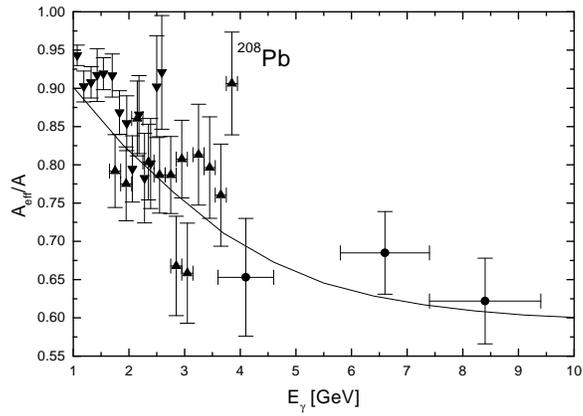,width=9.5cm}}
\caption{Effective mass number as function of photon energy for
photoabsorption on $^{208}$Pb. See Fig.~\ref{shad_12c} for references to the
experimental data.}
\label{shad_208pb}
\end{figure}

 In Figs.~\ref{shad_12c} and \ref{shad_208pb}
the resulting shadowing effect is
compared to experimental data as function of photon energy for $^{12}$C and
$^{208}$Pb. Within the experimental error bars the agreement is quite
satisfactory.
\par We can define a shadowing factor $s_N(\vec{r})$ for an in-medium
photon-nucleon cross section:
\begin{equation}
s_N(\vec{r})=1-\frac{S(\vec{r})}{\sigma_{\gamma N}} \quad,
\end{equation}
so that we can write the total photon-nucleus cross section as integral over
in-medium shadowed single nucleon cross sections:
\begin{equation}
\sigma_{\gamma A}=\int d^3r \rho(\vec{r}) s_N(\vec{r}) \sigma_{\gamma N} \quad.
\end{equation}
\par
\begin{figure}[h]
\centerline{\psfig{figure=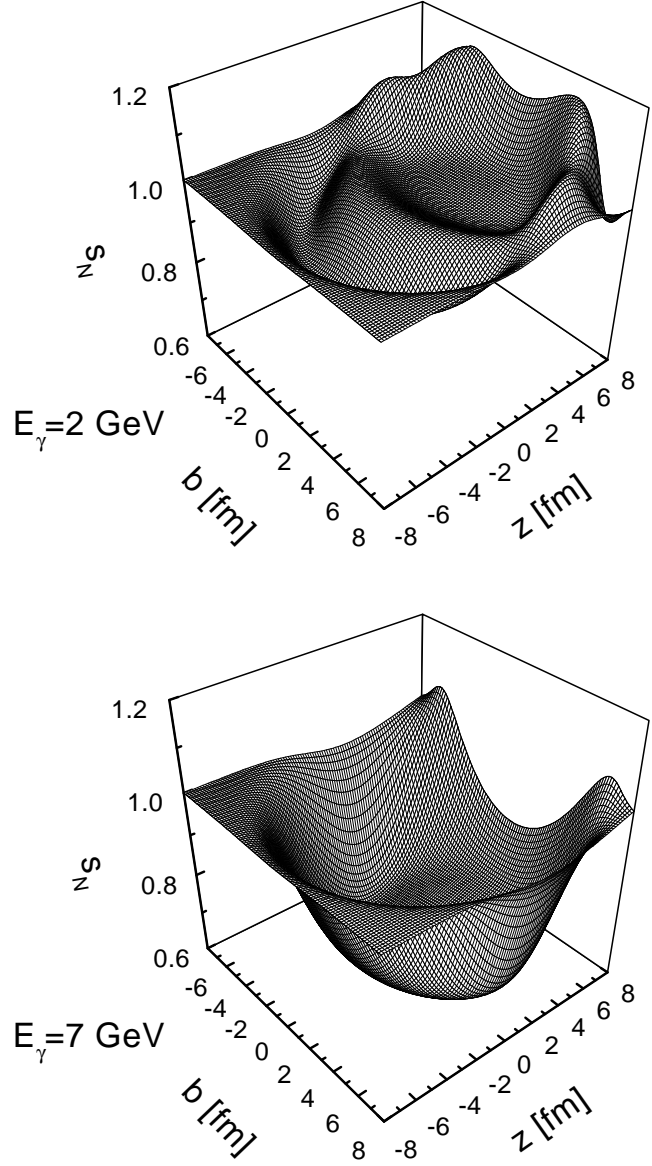,width=12cm}}
\caption{Coordinate dependence of the shadowing factor $s_N$ for a $\gamma$Pb
reaction at 2 GeV (upper part) and 7 GeV (lower part).}
\label{sn_pb_2_7}
\end{figure}

 In Fig.~\ref{sn_pb_2_7} we show the shadowing factor
$s_N(\vec{r})$ for $^{208}$Pb at photon energies of
2 and 7 GeV. At 2 GeV
there are some interference structures but since $s_N$ varies only between
0.8
and 1.2 the influence of shadowing on our calculations for particle
production is very small. At 7 GeV shadowing is much more important. Due to
interference effects one sees a rise of $s_N$ for larger $z$ that one naively
would not expect.
\par In our calculations the same shadowing factor $s_N(\vec{r})$ is used
for all
partial cross sections, so that, for example,
the primary in-medium cross section
$\sigma_{\gamma N \to N m}^{med}$ for photoproduction
of a meson m in $\gamma N \to N m$ is related to the vacuum cross section
$\sigma_{\gamma N \to N m}^{vac}$ via:
\begin{equation}
\sigma_{\gamma N \to N m}^{med}=
s_N(\vec{r})
\sigma_{\gamma N \to N m}^{vac}.
\end{equation}
The final state interactions are then treated in the transporttheoretical
framework described in Section~\ref{sec:buu}.
The initial state interactions of the incoming photon are thus described in
a model that contains quantum-mechanical coherence effects whereas the final
state interactions are treated in a completely
incoherent way.
\section{Results}
\label{sec:results}
\subsection{Photoproduction of pions and etas}
\label{sec:results1}
\begin{figure}[h]
\centerline{\psfig{figure=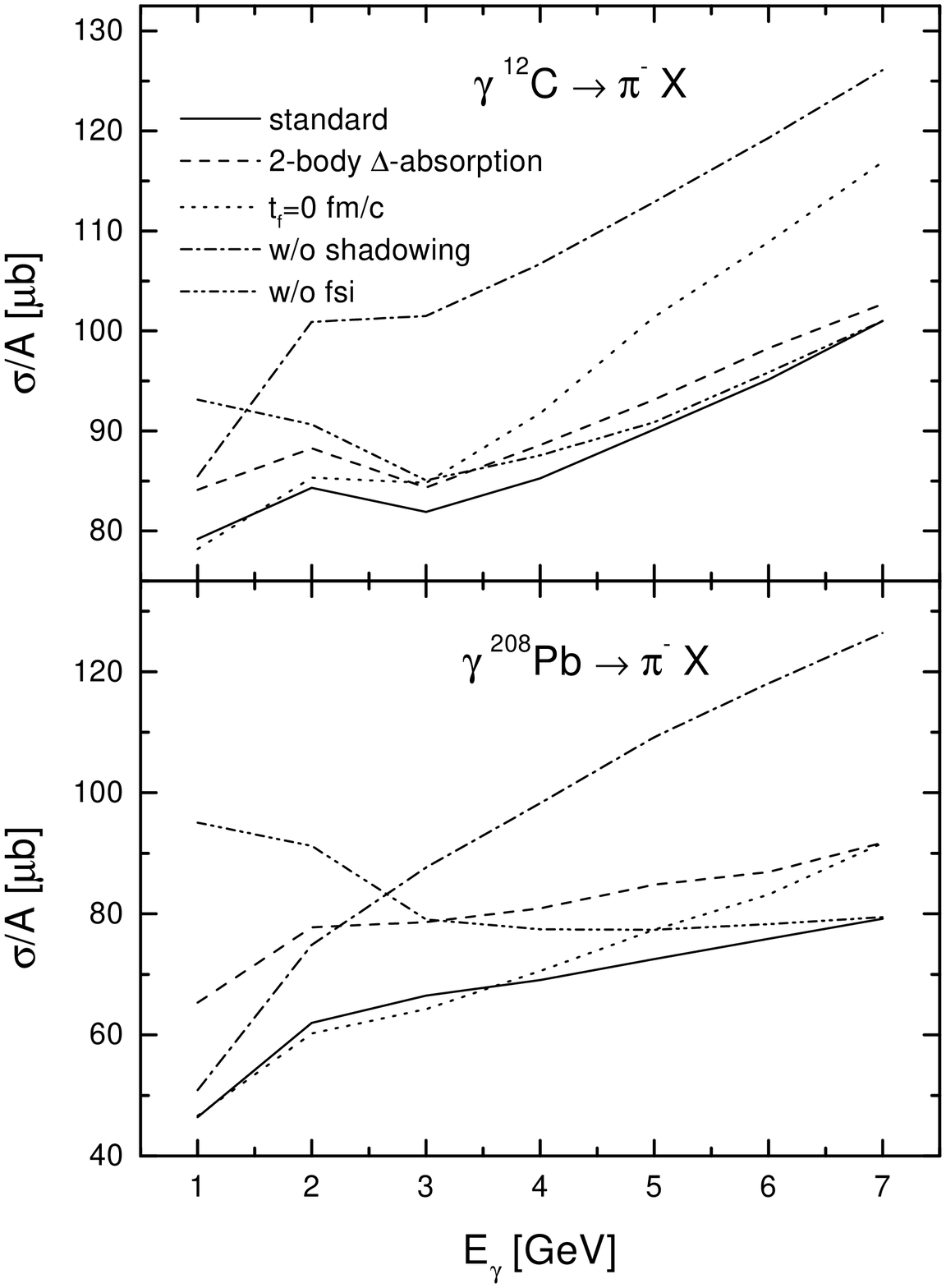,width=11cm}}
\caption{Total cross sections for $\pi^-$-production in $\gamma$C
(upper part) and
$\gamma$Pb (lower part) reactions. See text for a detailed
explanation of the different lines.}
\label{fig8}
\end{figure}

In Fig.~\ref{fig8} we present the results of our calculations for the total
$\pi^-$ production cross section in $^{12}$C (upper part) and $^{208}$Pb (lower
part). The solid lines display our 'standard' calculations.
One sees that the cross section per nucleon is larger for $^{12}$C because
nuclear shadowing and final state interaction are less effective.
For $^{12}$C there is also a local maximum around 2 GeV in the excitation
function that is not present in $^{208}$Pb. This maximum is mainly caused by
an interplay of the rising $\pi$-production cross section
and the
onset of shadowing. For $^{208}$Pb the decrease between 2 and 3 GeV due
to shadowing is neutralized by secondary collisions of the primary produced
particles that contribute to the total yield.
\par For the dashed lines we used a two-body absorption for the
$\Delta(1232)$-resonance only instead of the absorption introduced
in Ref.~\cite{lehr} that gives an enhanced $\Delta$-absorption. With
increasing photon energy the treatment of the $\Delta$-absorption becomes
less important. For $^{12}$C the improved, new treatment reduces the pion yield
by about 5\%, for $^{208}$Pb by about 20\%.
\par The use of a string fragmentation model in hadronic transport models
always requires the introduction of a finite 'formation time'. In our
calculations we use $t_f=0.8$ fm/c (in the rest frame of the produced
particles) which leads to a good reproduction of experimental data on pion
production
in heavy-ion collisions
at SPS energies. While this value lies in a reasonable range
our formation time prescription is nonetheless questionable as we neglect
any interaction of the 'strings' during their formation
with the surrounding nuclear medium.
From a practical point of view this prescription
should thus be regarded as a parameterization of our ignorance with respect to
the role of partonic degrees of freedom. In order to explore the resulting
uncertainties we have also performed calculations with $t_f=0$ (dotted lines
in Fig.~\ref{fig8}). This enhances the final state interactions of the
produced particles. Here one should note that by the final state interactions
the particle yield
is not only reduced but
can also be enhanced
when a primarily produced particle with high energy strikes
another nucleon, as e.g. in $\pi N \to \pi \pi N$.
From Fig.~\ref{fig8} one sees that the
formation time becomes more important with increasing photon energy. For
$^{12}$C the calculation with zero formation time gives an enhancement of the
pion yield at 7 GeV by about 15\%. This enhancement is less pronounced in
$^{208}$Pb because the absorption of the secondary produced particles
is more effective.
\par In Fig.~\ref{fig8} we also show results of calculations without
shadowing for the incoming photon (dash-dotted lines).
Because of the coordinate dependence of
the shadowing factor the effect of shadowing is in principle
not simply proportional to the effect seen in the total absorption
cross section. However, the deviations from this proportionality are quite
small as one sees by comparing the results in Fig.~\ref{fig8} to the
effective mass numbers in Figs.~\ref{shad_12c} and \ref{shad_208pb}.
\par The effect of the final state interaction of the produced particles
is, as
expected, rather different for $^{12}$C and $^{208}$Pb. In Fig.~\ref{fig8}
calculations without final state interaction are displayed by the
dot-dot-dashed lines. Now there are only two competing mechanisms that
influence the shape of these excitation functions. On the one hand
the particle yield in elementary photon-nucloen collisions increases
with photon energy. On the other hand the shadowing effect becomes also more
important for higher energies. In the case of $^{208}$Pb the shadowing
effect dominates and the cross section decreases monotonously with increasing
photon energy. For $^{12}$C there is first a decrease of the pion yield
up to a photon energy of 3 GeV and then an increase because shadowing is
here less
important than for $^{208}$Pb.
\par
\begin{figure}[h]
\centerline{\psfig{figure=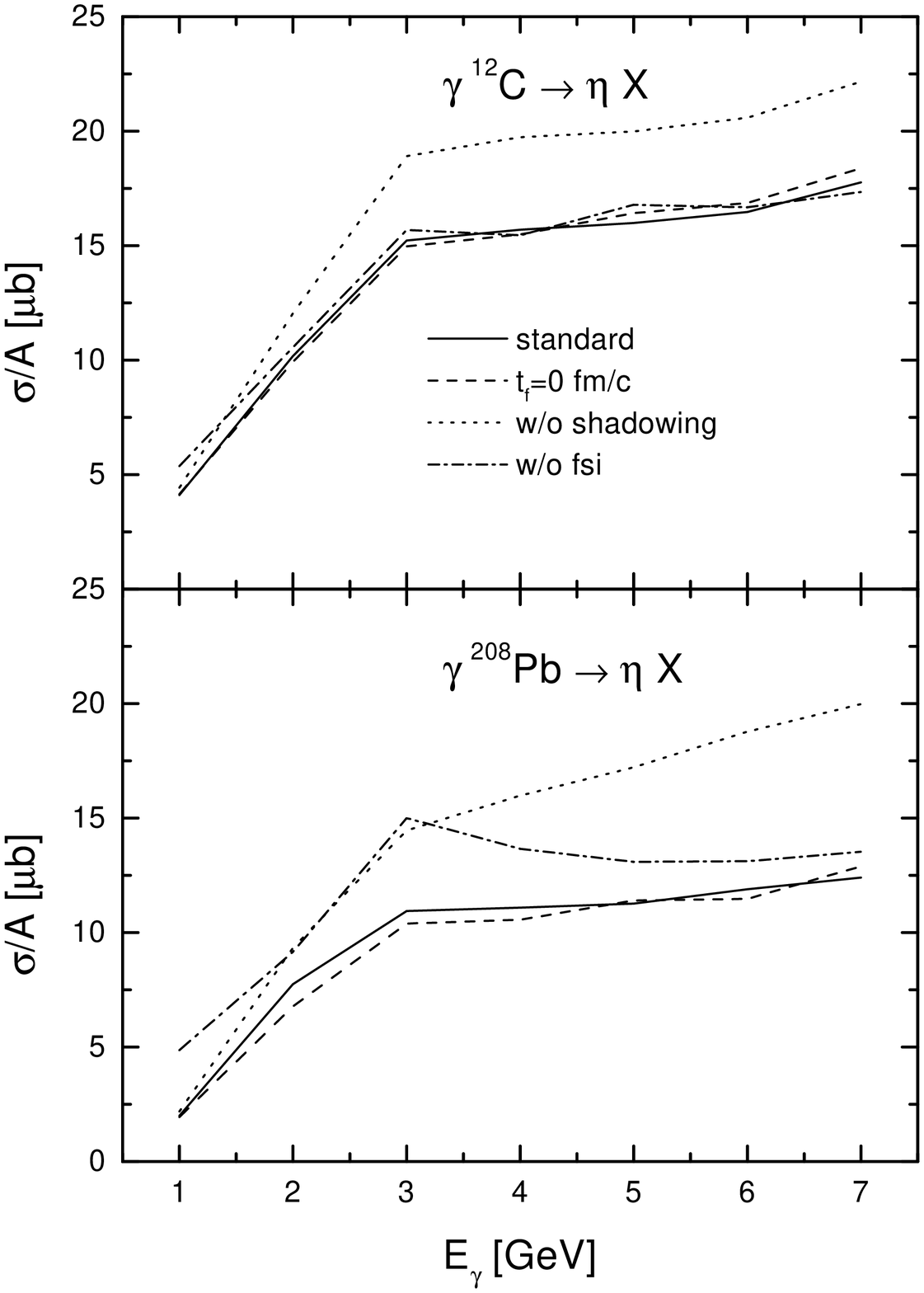,width=11cm}}
\caption{Total cross sections for $\eta$-production in $\gamma$C
(upper part) and
$\gamma$Pb (lower part) reactions. See text for a detailed
explanation of the different lines.}
\label{fig9}
\end{figure}

 In Fig.~\ref{fig9} we show the effects of the same scenarios on
$\eta$ photoproduction. First one observes a strong increase of the
total cross section from 1 to 3 GeV by about a factor of 4
which is simply due to the opening of phase space.
The effect of the
formation time is again determined by an interplay between secondary production
and absorption which results in a very small net effect.
\par The shadowing effect is, as expected, very similar to the case of pion
production. For $^{12}$C the calculation without final state interaction
gives practically the same result as the 'standard' calculation. This is
quite different for $^{208}$Pb where the final state interaction reduces the
yield, in particular for low photon energies, significantly.
\par
\begin{figure}[h]
\centerline{\psfig{figure=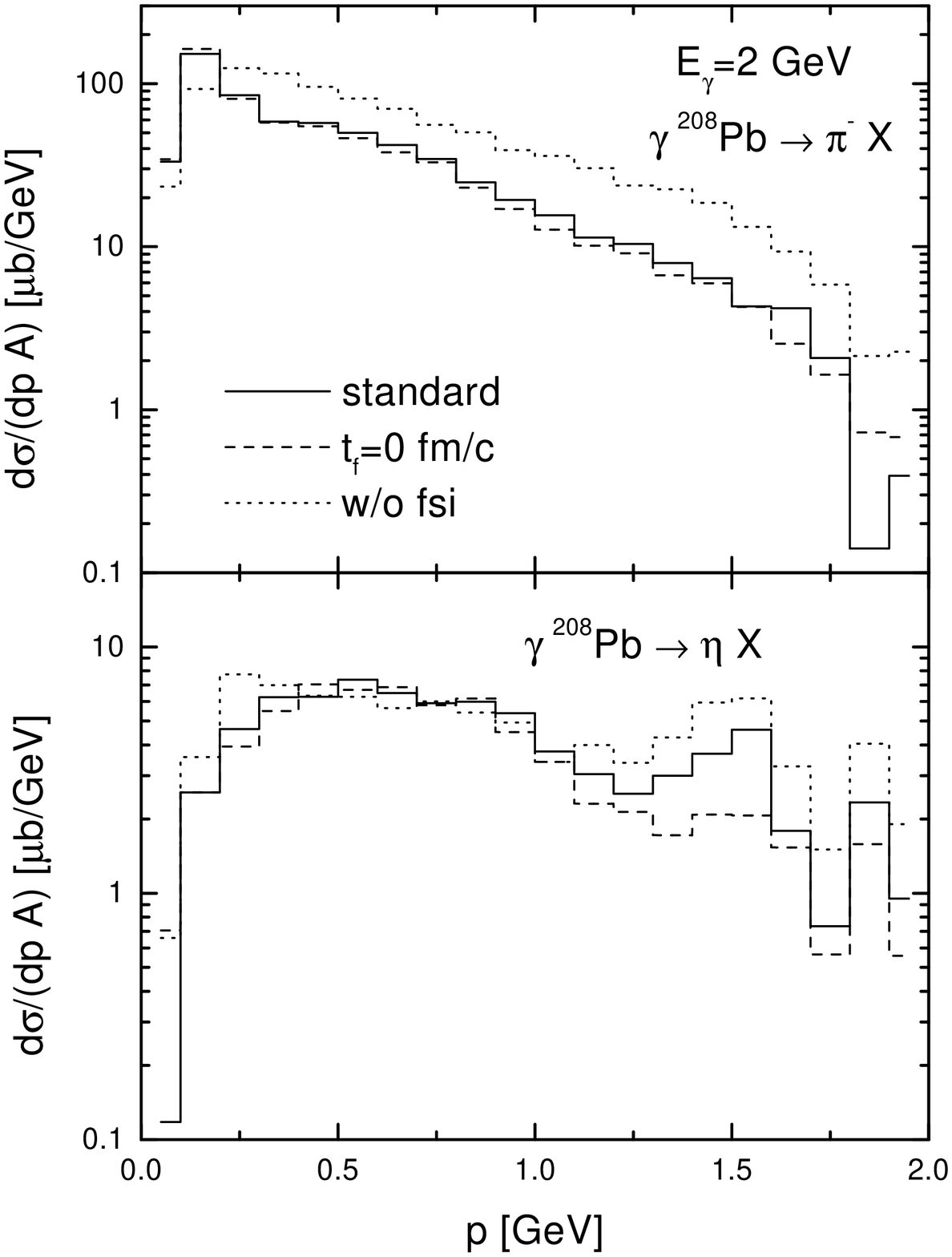,width=11cm}}
\caption{Momentum differential cross sections for $\pi^-$ (upper part) and
$\eta$ (lower part) production in $\gamma$Pb reactions at a photon energy
of $E_\gamma=2$ GeV.}
\label{fig92}
\end{figure}

\begin{figure}[h]
\centerline{\psfig{figure=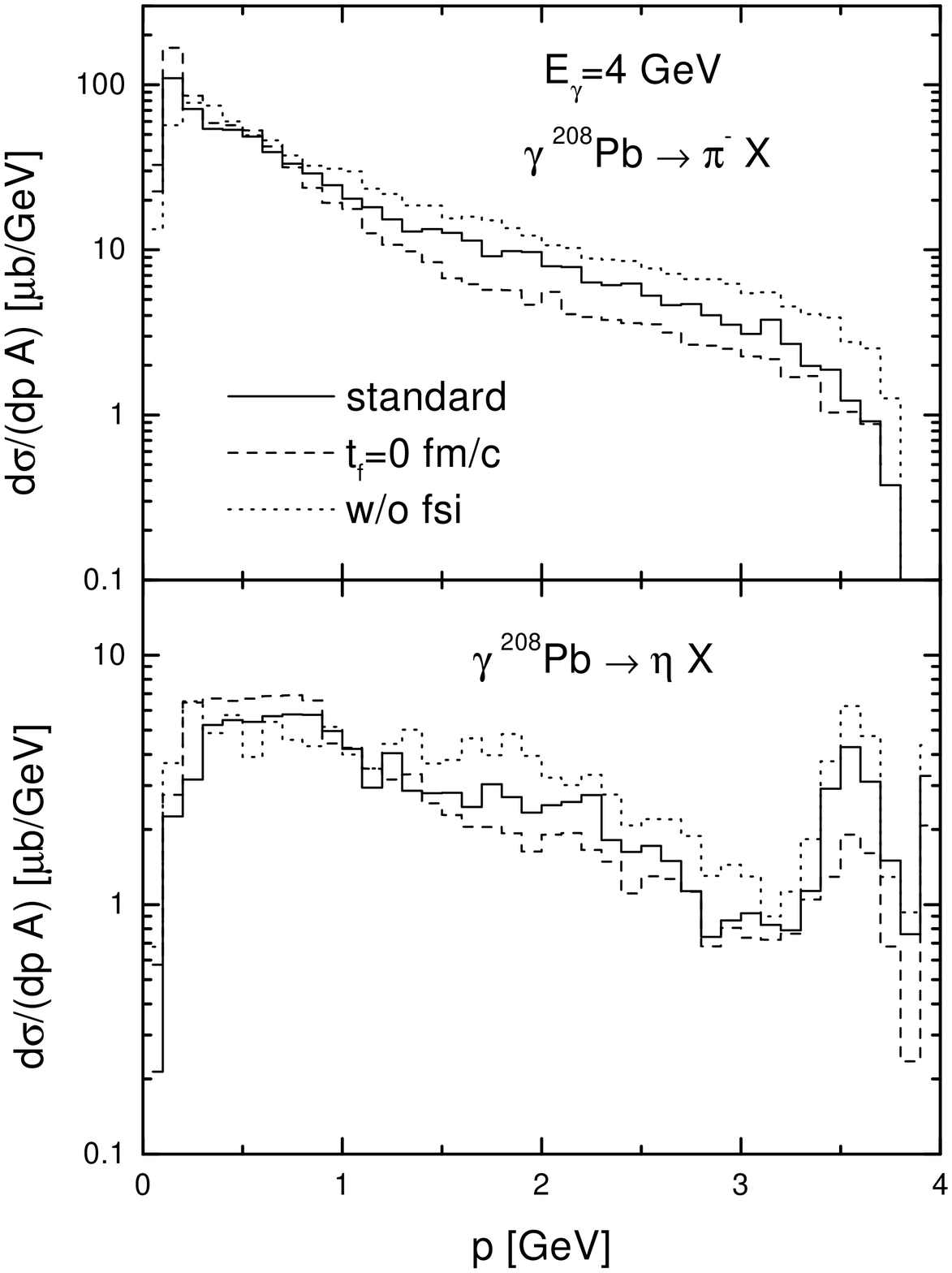,width=11cm}}
\caption{Momentum differential cross sections for $\pi^-$ (upper part) and
$\eta$ (lower part) production in $\gamma$Pb reactions at a photon energy
of $E_\gamma=4$ GeV.}
\label{fig91}
\end{figure}

 Since the total meson production yield is, as discussed above, determined
by different effects that partly cancel each other it is instructive to look
at more exclusive observables.
In Figs.~\ref{fig92} and \ref{fig91} we therefore present
momentum differential cross sections for the production of pions (upper part)
and etas (lower part) at
photon energies of 2 and 4 GeV in $^{208}$Pb. Here we show our 'standard'
calculations (solid lines) as well as calculations without formation
time (i.e. maximum
final state interaction) (dashed lines) and without
final state interaction (dotted lines).
In the
pion case the spectrum is getting 'softer' with increasing final state
interactions.
\par The $\eta$-spectra show a pronounced structure at high momenta that
is caused from exclusive processes $\gamma N \to N \eta$ that are strongly
forward peaked. In $\pi^-$ production such a structure is not present because
the string fragmentation model used here does not include flavor exchange
mechanisms. We leave the inclusion of such processes for future work.
The influence of the
final state interaction on the $\eta$-spectrum is similar to the pion case:
the final state interactions mainly shift the spectrum to lower energies.
\par
\begin{figure}[h]
\centerline{\psfig{figure=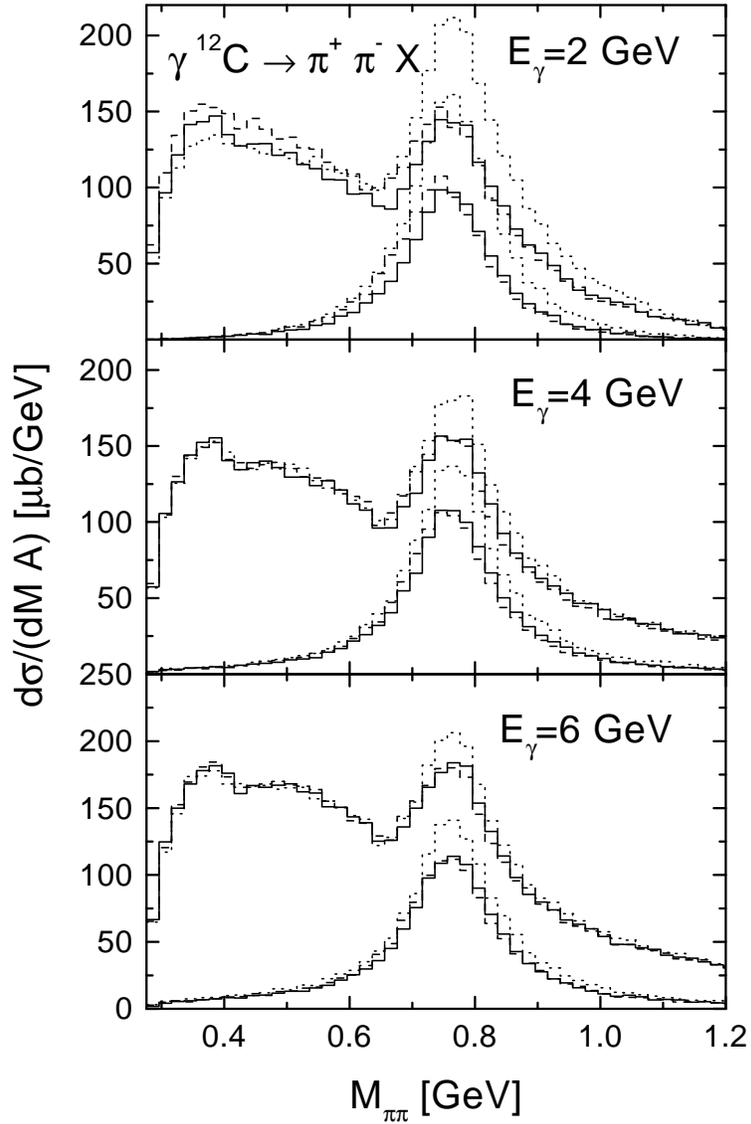,width=11cm}}
\caption{
Invariant mass spectrum of inclusive $\pi^+ \pi^-$ production in $\gamma$C
reactions. The dashed lines are the results with a dropping mass scenario
for the $\rho$ meson. The dotted lines result without final state
interactions. The total mass differential cross sections as well as the
contributions coming from the $\rho$ meson are displayed.
}
\label{c_rho}
\end{figure}

\begin{figure}[h]
\centerline{\psfig{figure=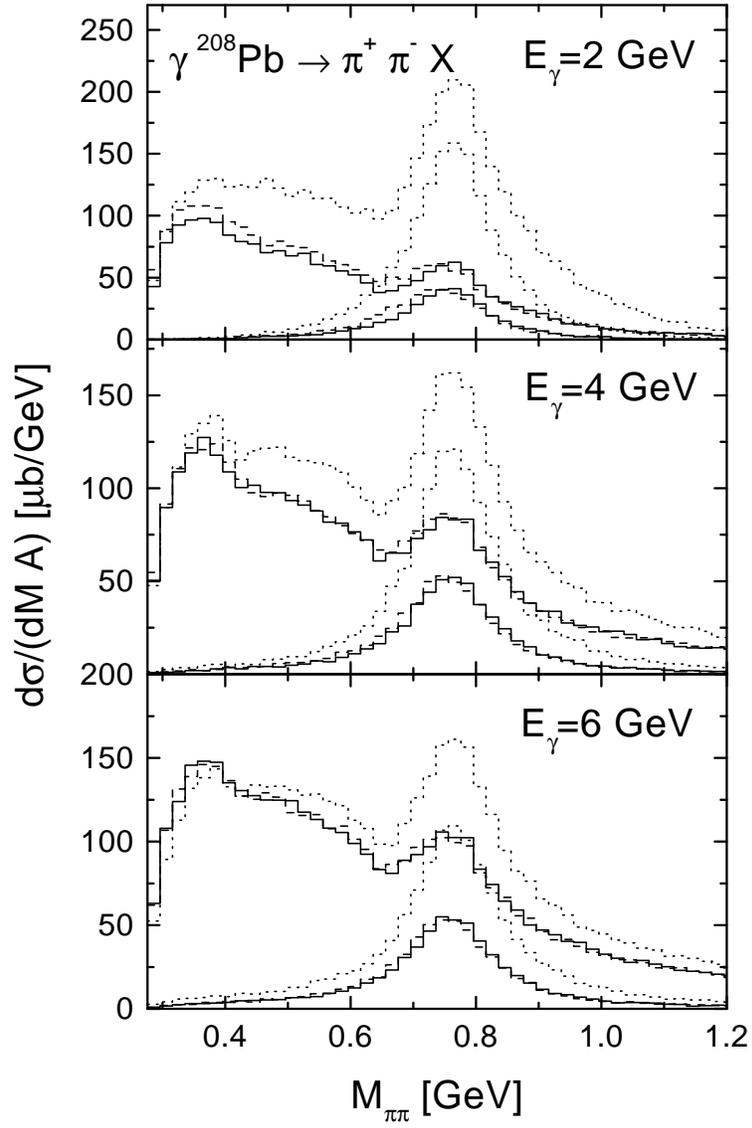,width=11cm}}
\caption{
Same as Fig.~\ref{c_rho} for $\gamma$Pb reactions.}
\label{pb_rho}
\end{figure}

 In Figs.~\ref{c_rho} and \ref{pb_rho} we present the results of our
calculations for the $\pi^+ \pi^-$ invariant mass spectra at photon energies
of 2, 4, and 6 GeV in $^{12}$C and $^{208}$Pb, respectively. We show at each
case
the total mass differential cross section as well as the contribution
coming from $\rho^0$ decays. The effect of the final state interaction is,
as expected, much larger for $^{208}$Pb than for $^{12}$C. While the peak
of the $\rho$ meson clearly dominates the spectrum in $^{12}$C it is,
especially for
low photon energies, harder to identify in $^{208}$Pb.
\par In Ref.~\cite{dilgam} we have investigated the observable effects of
medium modifications of the vector mesons $\rho$ and $\omega$ through their
$e^+ e^-$ decay in photoproduction at energies between 0.8 and 2.2 GeV. In
this study we have found an enhancement of the yield of intermediate mass
($\sim 500$ MeV) dileptons by about a factor 3 when the mass of the
$\rho$ meson was reduced in the nuclear medium according to the predictions of
Refs.~\cite{brown,hatsuda}:
\begin{equation}
\mu^*=\mu-0.18 m_\rho^0 \frac{\rho(\vec{r})}{\rho_0},
\end{equation}
where $m_\rho^0$ denotes the pole mass of the $\rho$ meson.
In Figs.~\ref{c_rho} and \ref{pb_rho} the dashed lines show the calculations
with such a dropping mass scenario. One sees that the $\pi^+ \pi^-$ spectrum
is hardly influenced by such a medium modification. This is simply due to
the fact that the pions have a very short mean free path in the nuclear
medium ($\sim 1$ fm). Therefore the probability that two pions which stem
from a decay of a $\rho$ meson at a relevant density are both able to
propagate to the vacuum without rescattering is very low.
\subsection{Photoproduction of kaons and antikaons}
\label{sec:results2}
\begin{figure}[h]
\centerline{\psfig{figure=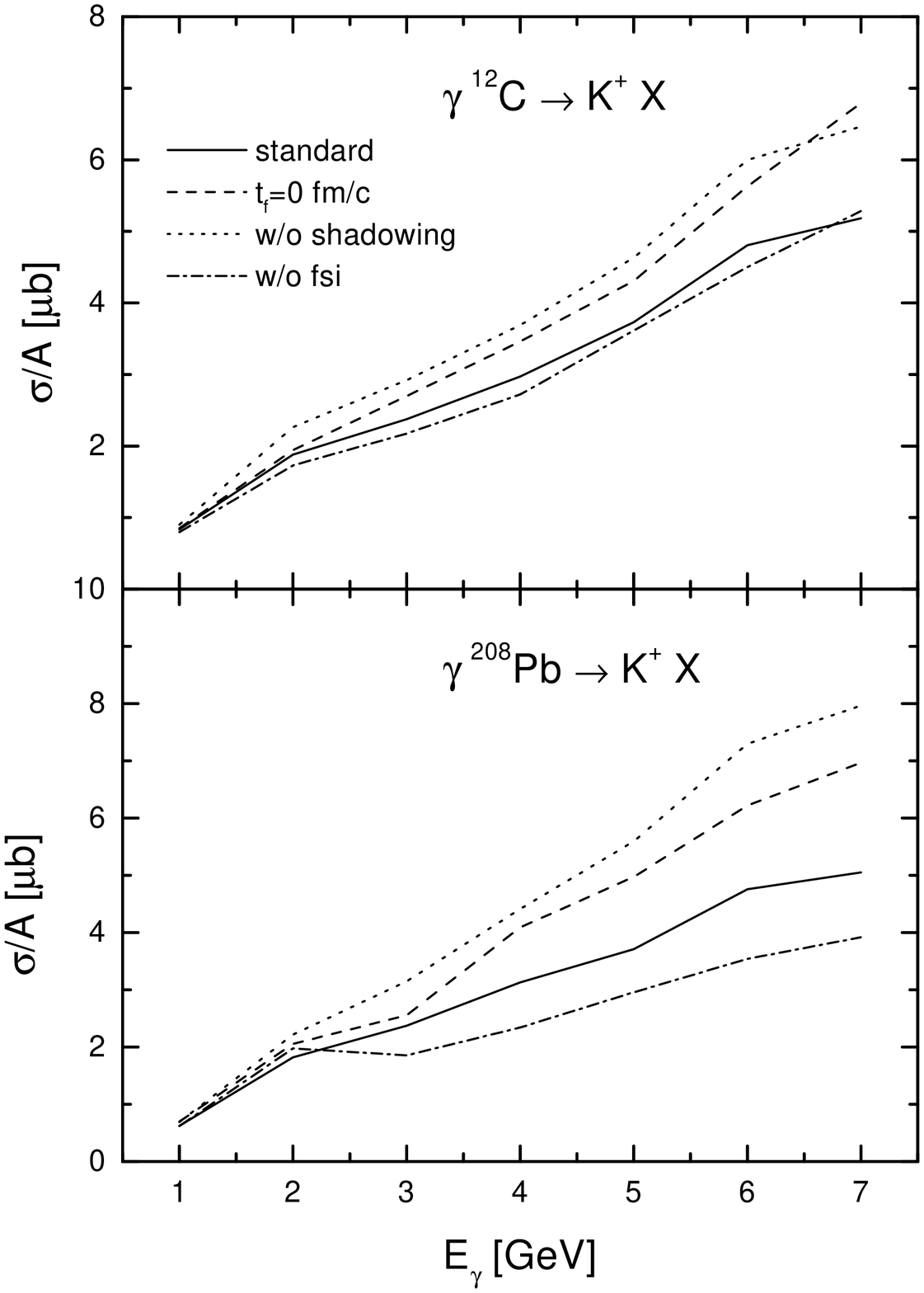,width=11cm}}
\caption{Total cross sections for $K^+$-production in $\gamma$C
(upper part) and
$\gamma$Pb (lower part) reactions.}
\label{fig10}
\end{figure}

In Fig.~\ref{fig10} we present our results for $K^+$-production in
$^{12}$C and
$^{208}$Pb. We again show, like for pion and eta production,
the results of different model calculations: a 'standard' calculation
(solid lines) that includes all effects, a calculation with zero formation
time (dashed lines), without shadowing (dotted lines), and without final
state interaction (dash-dotted lines). The primary produced
$\bar{s}$-quarks can not be annihilated in the nuclear medium and all of them
are finally contained in $K^+$ and $K^0$ mesons. Therefore the final state
interaction can only enhance the $K^+$ yield. From Fig.~\ref{fig10} one sees
that for $^{12}$C the number of secondary produced $K^+$-mesons is almost
negligible while for $^{208}$Pb they amount to about 30\% of the total yield.
In a calculation with zero formation time the cross section for $K^+$
production is enhanced for large photon energies in $^{12}$C by about 20\%
and in $^{208}$Pb by about 40\%. This enhancement is mainly caused
by collisions of primary high energy pions with nucleons.
Such pions have a large formation
time $t_f^{lab}=\gamma t_f$ in the lab system, where $\gamma=E_\pi/m_\pi$ is
a Lorentz factor, which suppresses secondary interactions.
\par The shadowing effect is very similar to the one for pion or
eta production. The fact that the calculations without shadowing and zero
formation time are practically identical for both nuclei is accidental.
\par
\begin{figure}[h]
\centerline{\psfig{figure=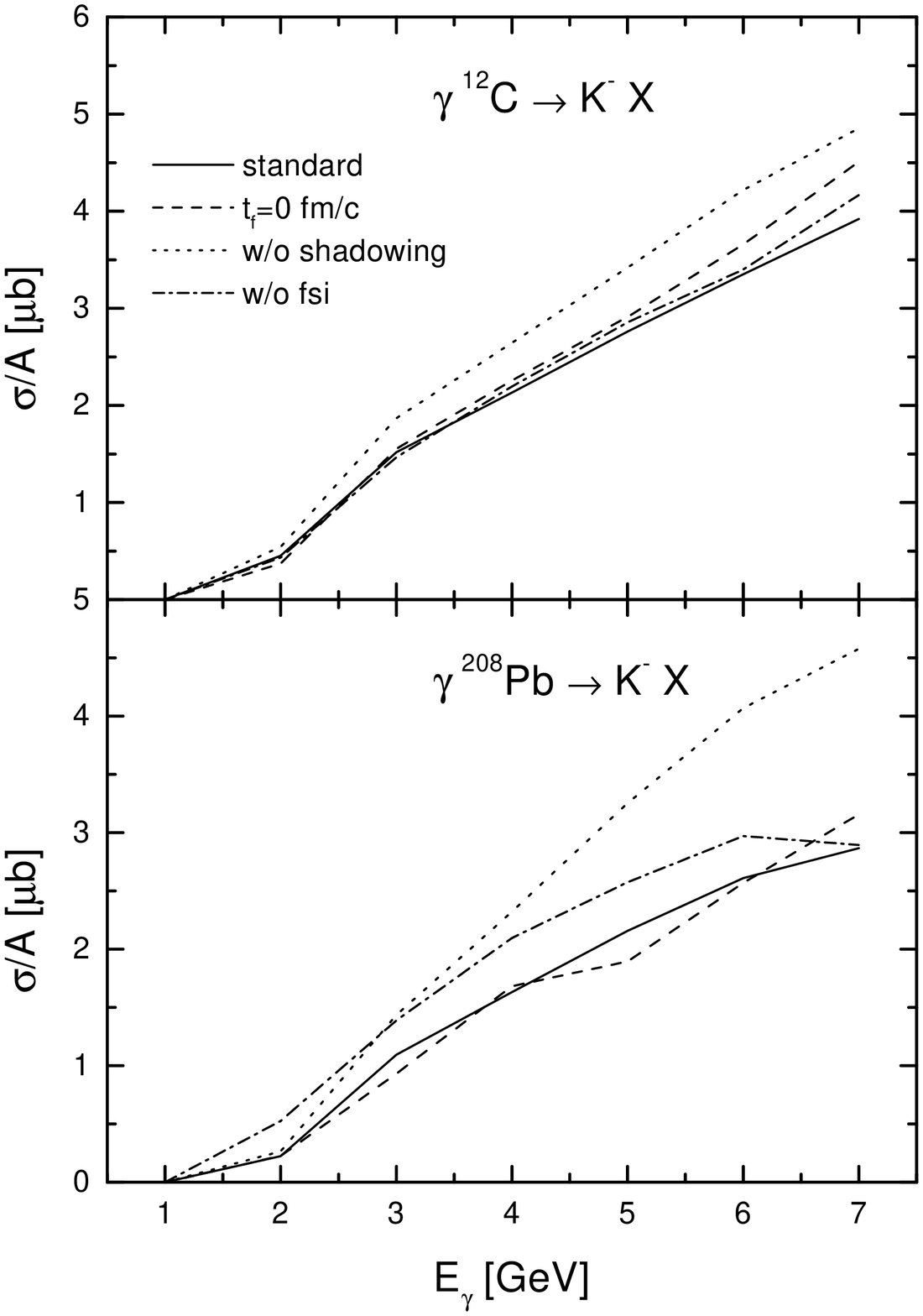,width=11cm}}
\caption{Total cross sections for $K^-$-production in $\gamma$C
(upper part) and
$\gamma$Pb (lower part) reactions.}
\label{fig11}
\end{figure}

 In Fig.~\ref{fig11} we show our results for $K^-$-production. In contrast
to $K^+$-production primary $K^-$-mesons can be absorbed via processes
$K^- N \to Y \pi$. For $^{12}$C absorption and secondary production nearly
cancel each other as can be seen by comparing the 'standard' calculation
(solid line) and the calculation without final state interaction
(dash-dotted line). The calcultion with zero formation time gives a slightly
larger result. In $^{208}$Pb the absorption mechanism is a little more
dominant. Therfore, the calculation without final state interaction gives
the largest result.
\par By comparing the total $K^+$- and $K^-$-production in Figs.~\ref{fig10}
and \ref{fig11} one sees that in our calculation almost as many antikaons
as kaons are produced. This is due to the fact that the string fragmentation
model produces much more antikaons than hyperons in a
$\rho^0$-nucleon collision which might be an artifact of the model. However,
since there are presently no experimental data for inclusive antikaon
production in photon-nucleon collisions available it is difficult to
check this point.
Therefore we stress here that
our results for photoproduction in nuclei are only valid under the caveat that
our description of the elementary photon-nucleon collision is correct.
\par
\begin{figure}[h]
\centerline{\psfig{figure=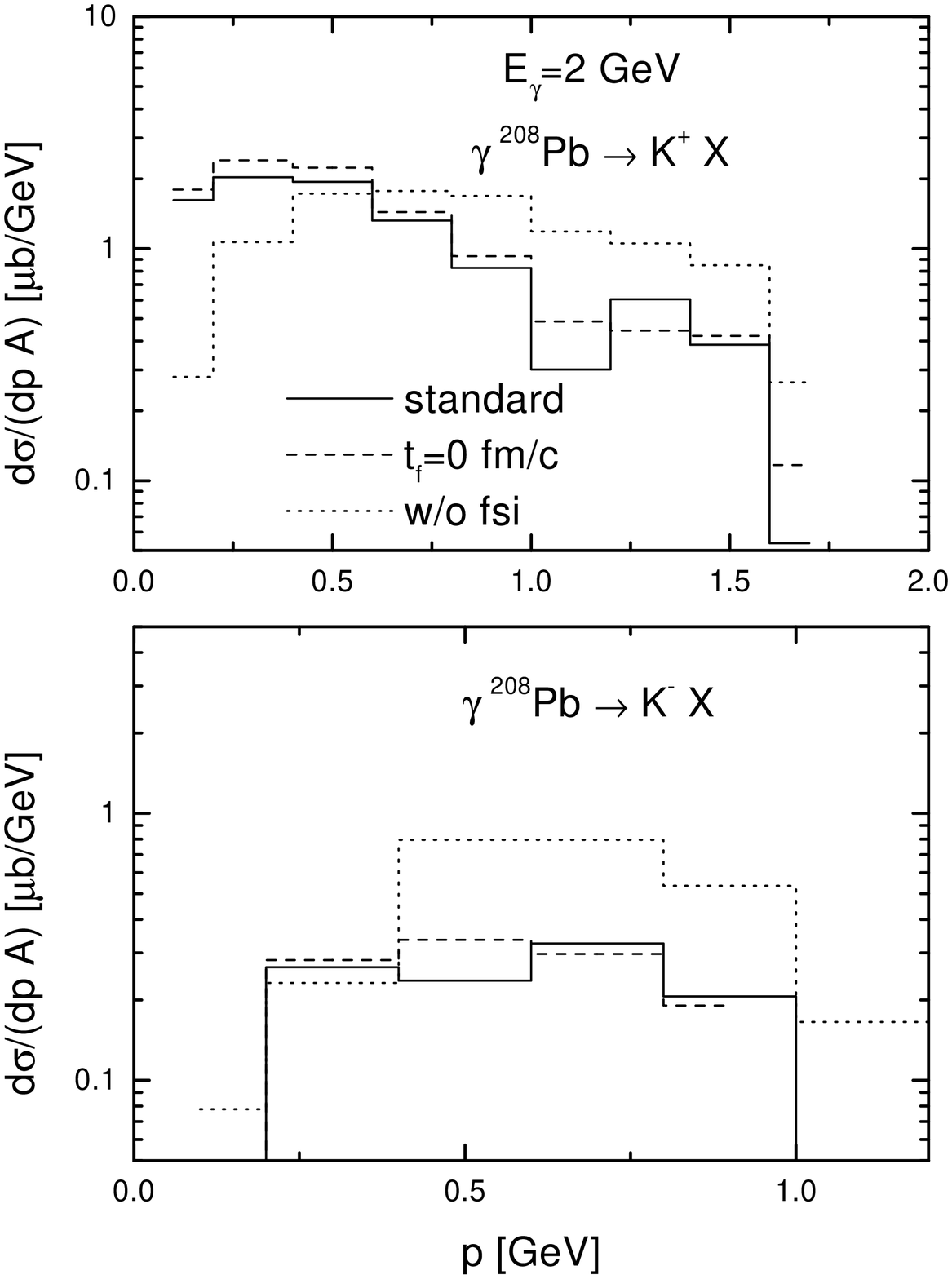,width=11cm}}
\caption{Momentum differential cross sections for $K^+$ (upper part) and
$K^-$ (lower part) production in $\gamma$Pb reactions at a photon energy
of $E_\gamma=2$ GeV.}
\label{fig112}
\end{figure}

\begin{figure}[h]
\centerline{\psfig{figure=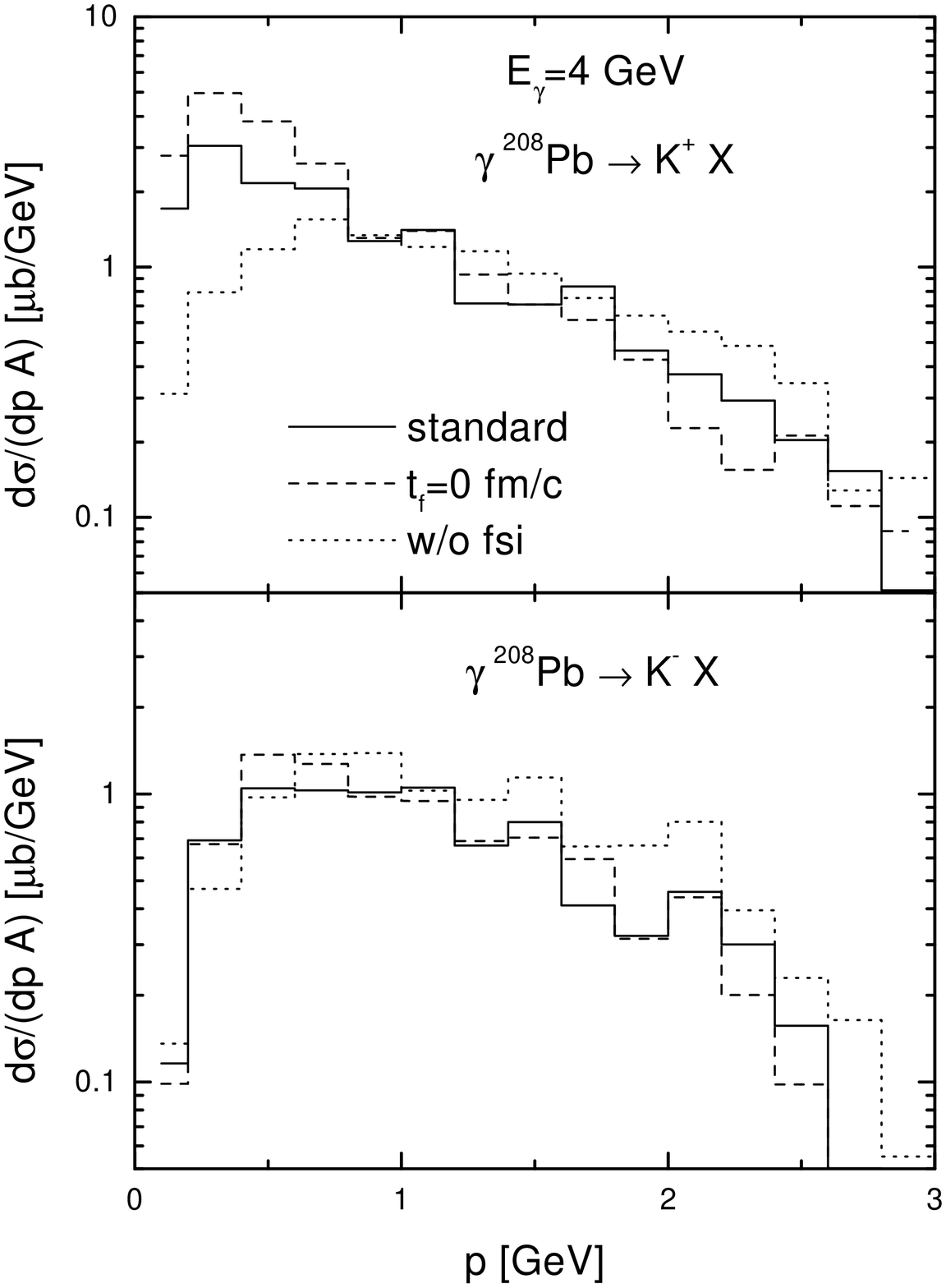,width=11cm}}
\caption{Momentum differential cross sections for $K^+$ (upper part) and
$K^-$ (lower part) production in $\gamma$Pb reactions at a photon energy
of $E_\gamma=4$ GeV.}
\label{fig111}
\end{figure}

 In Figs.~\ref{fig112} and \ref{fig111} we show momentum
differential cross sections for
production of $K^+$- and $K^-$-mesons in $^{208}$Pb at photon energies of
2 and
4 GeV. The solid lines are our 'standard' calculations, the dashed line are the
calculations with zero formation time, and the dotted lines display the result
without final state interaction. From the $K^+$-spectra one sees that the
final state interactions primarily enhance the low momentum yield.
For the $K^-$-mesons
the high momentum tail is significantly reduced by the final state
interactions. A comparison of these spectra to future experimental
data will therefore be helpful with respect to our understanding of the
interactions of kaons with nucleons in the nuclear medium.

\section{Summary}
\label{sec:sum}
We have presented a calculation of photoproduction of pions, etas, kaons, and
antikaons in the energy range from 1 to 7 GeV in $^{12}$C and $^{208}$Pb
within a
model that contains shadowing of the incoming photon and treats the outgoing
particles in a coupled channel
semi-classical BUU transport model.
It thus goes beyond the standard Glauber treatment because it can describe
all possible chains that can lead to the final state under investigation.
Predictions for total
cross sections as well as momentum differential cross sections were given.
\par We have investigated in detail the influence of shadowing for the
incoming photon and final state interaction of the produced particles
on our results. A comparison of experimental data to our results will
clarify if our treatment of the in-medium meson-nucleon interactions is
correct.
\par In particular, we have shown that $\pi^+ \pi^-$ invariant mass
spectra exhibit only
a very small sensitivity on medium modifications of the $\rho$ meson.
\par Our calculations of photoproduction of mesons in nuclei are based on the
input of photoproduction of mesons on nucleons for which so far only
few experimental data in the relevant energy range exist. New
experimental data are urgently needed in order to remove uncertainties
coming from the treatment of the elementary process. There are also
uncertainties coming from our description of the hadronic interactions
because not all channels have yet experimentally been measured or are
in principal unmeasurable. However, we do not expect these
uncertainties to have a large influence on the results reported here
since we have calculated rather inclusive observables for which the
total employed cross sections are most decisive.
\par In assessing the overall reliability of the predictions made here
we point out that the method (and code) used here has been shown to
provide a very good description of $\pi$, $2\pi$ and $\eta$
photoproduction data in the MAMI energy regime (up to 800 MeV)
\cite{prodpaper,lehr}. In the present paper we show that the method also
describes shadowing very well up to photon energies of a few GeV. The
same method also gives an excellent description of meson production in
heavy-ion collisions in the GeV regime, typically describing these data
within a factor $< 2$. In these latter reactions the same final state
interactions take place as in the present calculation. Thus, both the
initial state (shadowing) and the final state interactions with their
complex coupled channel effects are well under control. We thus feel
confident that the accuracy of the present calculations is as good as
that observed in the other reaction channels. 
\par Finally, we wish to mention that the
calculations reported here can be extended to the case of electroproduction
\cite{lehr}.

\begin{appendix}
\section{Kaon-nucleon and antikaon-nucleon collisions}
\label{app:kaon}
\begin{figure}[h]
\centerline{\psfig{figure=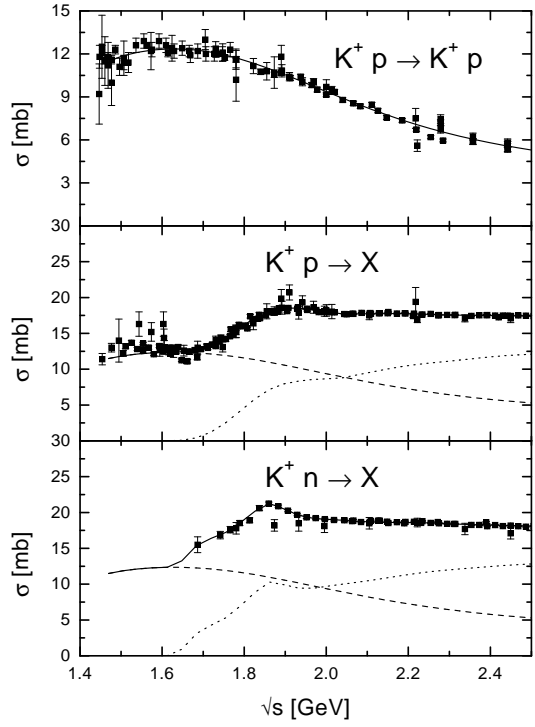,width=8cm}}
\caption{Parameterizations of the elastic $K^+ p$ scattering cross section
(upper part), the total $K^+ p$ cross section (middle part), and the total
$K^+ n$ cross section (lower part) as compared to the experimental data
from \protect\cite{landolt}. In the plots of the total cross sections the
dashed and the dotted lines show the elastic and the inelastic contributions,
respectively.}
\label{fig1}
\end{figure}

The elastic cross section $K^+ p \to K^+ p$ is parameterized for
invariant energies below the string threshold $\sqrt{s}<2.2$ GeV by the
following expression:
\begin{equation}
\sigma_{K^+ p \to K^+ p}=\frac{a_0 + a_1 p + a_2 p^2}{1+a_3 p + a_4 p^2} \quad,
\end{equation}
where $p$ denotes the kaon momentum in the rest frame of the nucleon and we
use: $a_0=10.508$ mb, $a_1=-3.716$ mb/GeV, $a_2=1.845$ mb/GeV$^2$,
$a_3=-0.764$/GeV, $a_4=0.508$/GeV$^2$. In Fig.~\ref{fig1} (upper part) we
show that this gives a good description of the experimental data. For
scattering on a neutron there exist in the relevant energy range
only a few experimental data for the
charge exchange process $K^+ n \to K^0 p$ \cite{glasser}
and no data for the process $K^+ n \to K^+ n$. Therefore, we assume on the
neutron the same total elastic cross section (including charge exchange):
\[
\sigma_{K^+ n \to K^+ n}=\sigma_{K^+ n \to K^0 p}=
\frac{1}{2} \sigma_{K^+ p \to K^+ p}.
\]
This is a rather crude approximation which is also not in line with the
experimental data for $K^+ n \to K^0 p$ at low momenta \cite{glasser}.
However, for the calculations presented here this is not essential since
all these cross sections are small and, in particular, for low momenta they
play only a small role for the phase space distributions of the particles
involved.
\par The inelastic kaon-nucleon cross section is obtained by a spline
interpolation
through selected data points of the total cross section after subtraction of
the elastic contribution. The resulting cross sections are displayed in
Fig.~\ref{fig1} (middle part for $K^+ p$, lower part for $K^+ n$).
We assume the inelastic cross section to consist only of $K \pi N$ states which
is a good approximation since these cross sections are only used for invariant
collision energies below 2.2 GeV.
The cross
sections for $K^0$ scattering on nucleons follow from isospin symmetry:
\[
\sigma_{K^0 p}=\sigma_{K^+ n},\quad \sigma_{K^0 n}=\sigma_{K^+ p}.
\]
\par
\begin{figure}[h]
\centerline{\psfig{figure=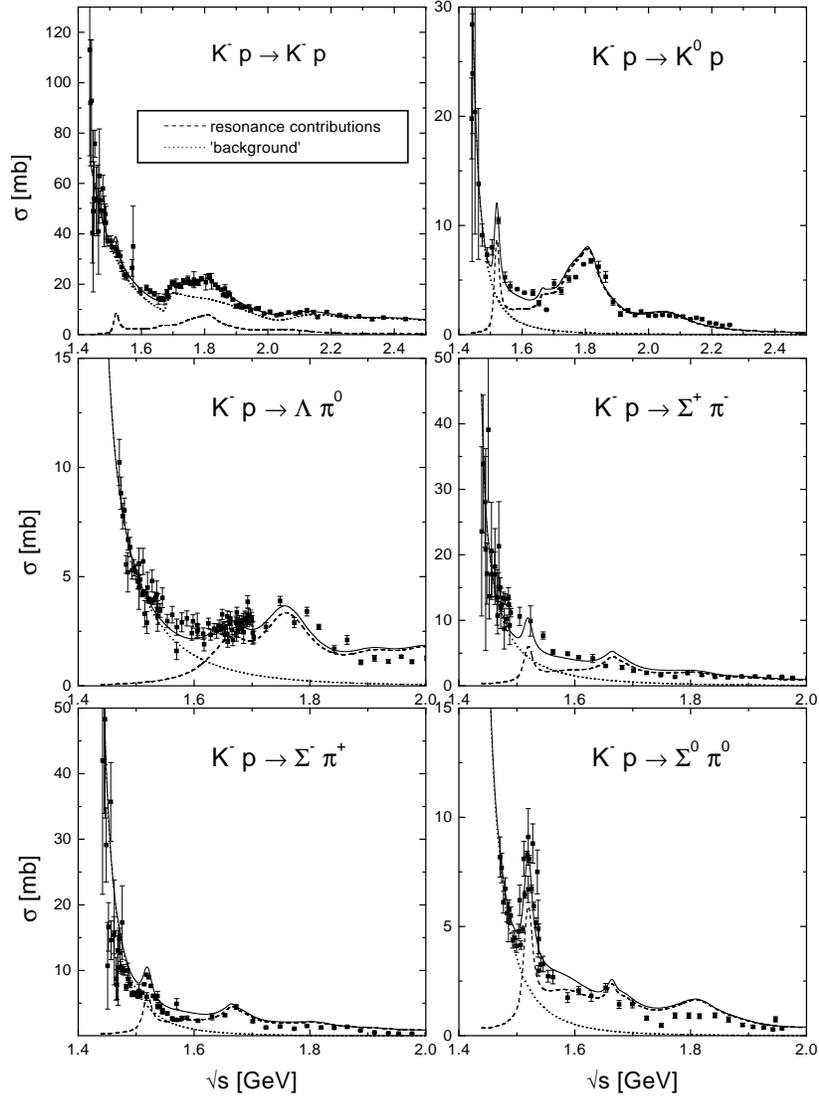,width=11cm}}
\caption{Cross sections for elastic scattering, charge
exchange, and hyperon production reactions in $K^- p$ collisions.
We show the total calculated cross sections (solid lines), the resonance
contributions (dashed lines), and the parameterizations of the
non-resonant background (dotted lines).
The experimental data are taken
from \protect\cite{landolt}.}
\label{fig2}
\end{figure}

 In case of antikaon scattering on nucleons we first get contributions
to the cross sections from the $S=-1$ resonances listed above which we treat
analogous to pion-nucleon scattering \cite{dilgam} as incoherent Breit-Wigner
type contributions. From Fig.~\ref{fig2} one sees that these contributions
alone -- unlike the case of pion-nucleon -- do not suffice to describe the
cross sections for the different channels. Therefore we have introduced
a non-resonant background cross section of the following form for the different
channels $i$ in $K^- p$ scattering:
\begin{equation}
\sigma^{bg}_{K^- p \to i}=a_0 \frac{p_f}{p_i s}
\left( \frac{a_1^2}{a_1^2+p_f^2} \right)^{a_2} \quad,
\label{kmpex}
\end{equation}
where $p_i$ and $p_f$ denote the cms momenta of the initial and final
particles, respectively and $\sqrt{s}$ is the total cms energy. The parameters
$a_j$ are given in Table~\ref{table2}. For $K^- p \to K^- p$ this
parameterization is only used for invariant energies below 1.7 GeV. For larger
energies we use a spline interpolation through selected experimental data
points because the broad bump in the experimental data around 1.8 GeV
can hardly be
described by a simple fit function. From Fig.~\ref{fig2} one sees that
our parameterizations give a good description of the experimental data
for $K^- p \to K^- p, K^0 n, \Lambda \pi^0, \Sigma^+ \pi^-, \Sigma^- \pi^+,
\Sigma^0 \pi^0$.
\par
\begin{figure}[h]
\centerline{\psfig{figure=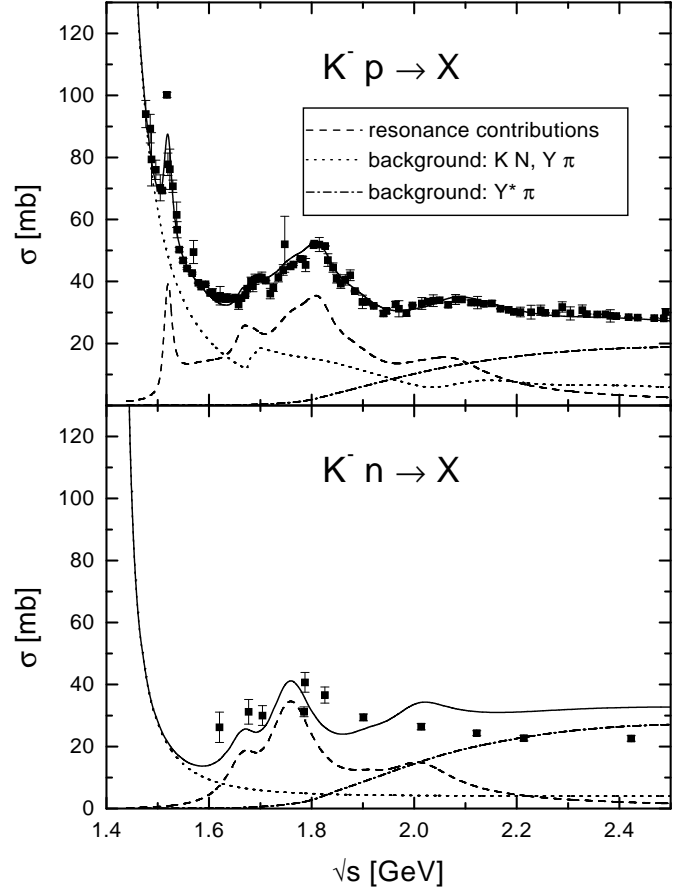,width=10cm}}
\caption{Total cross sections (solid lines)
for $K^-p$ (upper part) and $K^-n$ (lower
part) collisions and the different contributions: Resonance contributions
(dashed lines), non-resonant $\bar{K} N$ and $Y \pi$ contributions
(dotted lines), and $Y^* \pi$ contributions (dash-dotted lines).
The experimental data are taken
from \protect\cite{landolt}.}
\label{fig3}
\end{figure}

 In order to describe the total $K^- p$ cross section
in the relevant energy range it is necessary to also include channels with
more than 2 particles in the final state. This is done here by including
the process $\bar{K} N \to Y^* \pi$ with a constant matrix element for all
hyperon resonances. The cross section is then given as:
\begin{equation}
\sigma_{\bar{K} N \to Y^* \pi}=C \frac{\left|{\cal M}\right|^2}{p_i s}
\int^{\sqrt{s}-m_\pi} d\mu p_f {\cal A}_{Y^*}(\mu) \quad,
\label{ystarbg}
\end{equation}
where the isospin factor $C$ is given by the following expression of
Clebsch-Gordan coefficients:
\begin{equation}
C=\sum_I \left( \langle \frac{1}{2} \frac{1}{2} I_{z,\bar{K}} I_{z,N}|
\frac{1}{2} \frac{1}{2} I I_{z,tot} \rangle
\langle I_{Y^*} 1 I_{z,Y^*} I_{z,\pi}|
I_{Y^*}  1  I  I_{z,tot} \rangle \right)^2.
\end{equation}
In Eq.~(\ref{ystarbg}) ${\cal A}_{Y^*}$ stands for the spectral function of
the hyperon resonance $Y^*$ that is treated analogous to the ones for the
nucleonic resonances \cite{dilgam}. For the squared matrix element we use
$|{\cal M}|^2=22$ mb GeV$^2$ and we take for this process only the hyperon
resonances into account with a mass above 1.6 GeV. In Fig.~\ref{fig3}
(upper part) we
show that the sum of all partial contributions very well describes
the total $K^- p$ cross section.
\par For $K^- n$ scattering the elastic scattering cross section can be
described by the resonance contributions with a small constant background
cross section:
\[ \sigma^{bg}_{K^- n \to K^- n}=4 \ {\rm mb}. \]
The cross section for $K^- n \to \Lambda \pi^-$ follows from isospin
symmetry from $K^- p \to \Lambda \pi^0$:
\[
\sigma_{K^- n \to \Lambda \pi^-}=2 \sigma_{K^- p \to \Lambda \pi^0}.
\]
In the $\Sigma \pi$ channel we have from isospin symmetry:
\[ \sigma_{K^- n \to \Sigma^- \pi^0}=
\sigma_{K^- n \to \Sigma^0 \pi^-}
\]
and for the non-resonant background we assume:
\[
\sigma^{bg}_{K^- n \to \Sigma^- \pi^0}=
\frac{1}{2} \left( \sigma^{bg}_{K^- p \to \Sigma^- \pi^+}+
\sigma^{bg}_{K^- p \to \Sigma^+ \pi^-} \right).
\]
This is a rather crude approximation. However, for the calculations presented
here it is only of primary importance that the total cross section, summed
over all final states, are correctly described.
The result for the total $K^- n$ cross section is displayed in
Fig.~\ref{fig3} (lower part) and one sees that the available experimental
data are reasonably well reproduced. The cross sections for $\bar{K}^0$
scattering on nucleons immediately follow from isospin symmetry:
\[
\sigma_{\bar{K}^0 p}=\sigma_{K^- n},\quad \sigma_{\bar{K}^0 n}=\sigma_{K^- p}.
\]

\section{Strangeness production in photon-nucleon collisions}
\label{app:str}
\begin{figure}[h]
\centerline{\psfig{figure=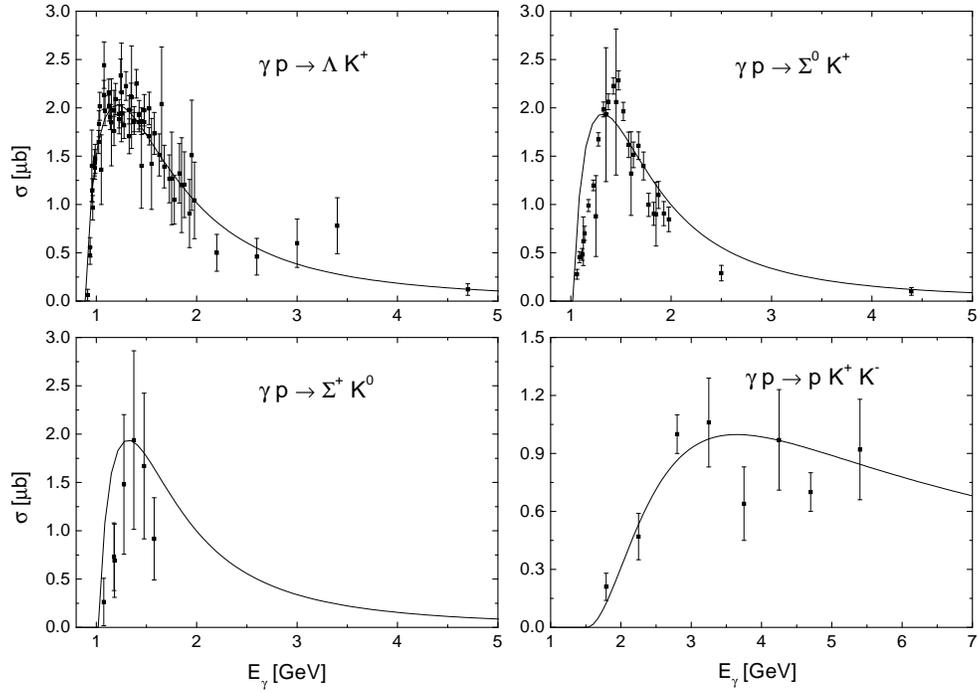,width=14cm}}
\caption{Parameterizations of the cross sections for
$\gamma p \to \Lambda K^+$,
$\gamma p \to \Sigma^0 K^+$,
$\gamma p \to \Sigma^+ K^0$,
and $\gamma p \to p K^+ K-$ compared to the experimental data from
Refs.~\protect\cite{landolt,saphir}.}
\label{fig31}
\end{figure}

The exclusive strangeness production processes
$\gamma N \to \Lambda K, \Sigma K, N K \bar{K}$ are fitted to available
experimental data. The channel
$YK$ is parameterized by the following expression:
\begin{equation}
\sigma_{\gamma N \to Y K}=a_0 \frac{p_f}{p_i s}
\frac{a_1^2}{a_1^2+(\sqrt{s}-\sqrt{s}_0)^2} \quad,
\end{equation}
where $\sqrt{s}_0$ denotes the invariant threshold energy. We assume the
constants $a_0$ and $a_1$ to be independent of the isospin states of the
incoming and outgoing particles and use:
$a_0^\Lambda=13$ $\mu$bGeV$^2$,
$a_0^\Sigma=15$ $\mu$bGeV$^2$,
$a_1^\Lambda=0.5$ GeV,
$a_1^\Sigma=0.4$ GeV.
In Fig.~\ref{fig31} it is shown that these parameterizations describe the
experimental data reasonably well.
\par The cross section for antikaon production $\gamma N \to N K \bar{K}$ is
parameterized by:
\begin{equation}
\sigma_{\gamma N \to N K \bar{K}}=a_0 \frac{16 (2 \pi)^7}{p_i \sqrt{s}}
\Phi_3
\frac{a_1^2}{a_1^2+(\sqrt{s}-\sqrt{s}_0)^2} \quad,
\end{equation}
where $\Phi_3$ denotes the 3-body phase space as, for example, given by
Eq.~(35.11) in Ref.~\cite{PDG}. From the experimental data for
$\gamma p \to p K^+ K^-$ we obtain $a_0=12$ $\mu$b and $a_1=0.7$ GeV
(resulting cross section shown in Fig.~\ref{fig31}). These
values are used for all isospin channels.
\par
\begin{figure}[h]
\centerline{\psfig{figure=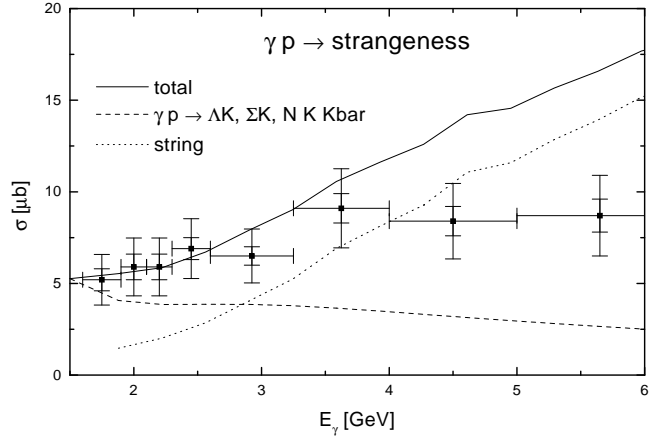,width=10cm}}
\caption{Total cross section for strangeness production in $\gamma p$
reactions. Experimental data are taken from
Ref.~\protect\cite{landolt}. The inner error bar displays the statistical
error, the outer one the sum of statistical and systematic error.}
\label{fig32}
\end{figure}

 In Fig.~\ref{fig32} we compare our treatment of strangeness production
in photon-proton reactions to the available experimental data on inclusive
strangeness production. The dashed line shows the sum of the exclusive
processes $\gamma p \to \Lambda K, \Sigma K, N K \bar{K}$; the dotted line
is the strangeness production cross section that results from the string
fragmentation model FRITIOF. We note that the string fragmentation model does
not produce the exclusive channels such that there is no double counting
involved here. The total inclusive strangeness production cross section, as
sum of the contributions from the exclusive channels and the string model, is
displayed as solid line in Fig.~\ref{fig32}.
The experimental data are described
rather well up to a photon energy of 4 GeV, but for larger energies the
experimental cross section is overestimated by about 50\%. The
experimental data seem to indicate a saturation of the cross section already
at 4 GeV which is hard to explain. Therefore we use the
described cross sections keeping in mind that out treatment of the
elementary photon-nucleon reaction might need to be refined when new and more
reliable experimental data will become available.
\end{appendix}

\newpage

\begin{table}[h]
\begin{center}
\begin{tabular}{|c|c|c||c|c|c|c|c|c|c|}
%\hline
 & & &\multicolumn{7}{|c|}{branching ratios [\%]} \\
\cline{4-10}
\raisebox{1.5ex}[-1.5ex]{Y($M_0$/MeV)} &
\raisebox{1.5ex}[-1.5ex]{$\Gamma_0$ [MeV]} &
\raisebox{1.5ex}[-1.5ex]{$J^P$} &
%\hline
%Y($M_0$/MeV) & $\Gamma_0$ [MeV] & $J^P$ &
$\Lambda \pi$ &
$N \bar{K}$ &
$\Sigma \pi$ &
$\Sigma^* \pi$ &
$\Lambda \eta$ &
$N \bar{K}^*$ &
$\Lambda^* \pi$ \\
\hline
$\Lambda(1116)$ & 0   & ${\frac{1}{2}}^+$ & 0 & 0  & 0 & 0 & 0 & 0&0 \\
\hline
$\Sigma(1189)$ & 0   & ${\frac{1}{2}}^+$ & 0& 0& 0& 0& 0&0 &0\\
\hline
$\Sigma(1385)$ & 36   & ${\frac{3}{2}}^+$ & 88 &0 & 12 &0 &0 &0&0 \\
\hline
$\Lambda(1405)$ & 50  & ${\frac{1}{2}}^-$ & 0 & 0  & 100 & 0 & 0 & 0&0 \\
\hline
$\Lambda(1520)$ & 16  & ${\frac{3}{2}}^-$ & 0 & 46  & 43 & 11 & 0 & 0&0 \\
\hline
$\Lambda(1600)$ & 150  & ${\frac{1}{2}}^+$ & 0 & 35  & 65 & 0 & 0 & 0&0 \\
\hline
$\Sigma(1660)$ & 100   & ${\frac{1}{2}}^+$ & 40& 20& 40& 0& 0&0 &0\\
\hline
$\Lambda(1670)$ & 35  & ${\frac{1}{2}}^-$ & 0 & 25  & 45 & 0 & 30 & 0&0 \\
\hline
$\Sigma(1670)$ & 60   & ${\frac{3}{2}}^-$ & 15& 15& 70& 0& 0&0 &0\\
\hline
$\Lambda(1690)$ & 60  & ${\frac{3}{2}}^+$ & 0 & 25  & 30 & 45 & 0 & 0&0 \\
\hline
$\Sigma(1750)$ & 90   & ${\frac{1}{2}}^-$ & 10& 30& 60& 0& 0&0 &0\\
\hline
$\Sigma(1775)$ & 120   & ${\frac{5}{2}}^-$ & 20& 45& 5& 10& 0&0 &20\\
\hline
$\Lambda(1800)$ & 300  & ${\frac{1}{2}}^-$ & 0 & 35  & 35 & 30 & 0 & 0&0 \\
\hline
$\Lambda(1810)$ & 150  & ${\frac{1}{2}}^+$ & 0 & 35  & 20 & 0 & 0 & 45&0 \\
\hline
$\Lambda(1820)$ & 80  & ${\frac{5}{2}}^+$ & 0 & 60  & 12 & 28 & 0 & 0&0 \\
\hline
$\Lambda(1830)$ & 95  & ${\frac{5}{2}}^-$ & 0 & 5  & 60 & 35 & 0 & 0&0 \\
\hline
$\Lambda(1890)$ & 100  & ${\frac{3}{2}}^+$ & 0 & 30  & 10 & 30 & 0 & 30&0 \\
\hline
$\Sigma(1915)$ & 120   & ${\frac{5}{2}}^+$ & 45& 10& 45& 0& 0&0 &0\\
\hline
$\Sigma(2030)$ & 180   & ${\frac{7}{2}}^+$ & 25& 25& 10& 15& 0&5 &20\\
\hline
$\Lambda(2100)$ & 200  & ${\frac{7}{2}}^-$ & 0 & 30  & 5 & 45 & 0 & 20&0 \\
\hline
$\Lambda(2110)$ & 200  & ${\frac{5}{2}}^+$ & 0 & 15  & 30 & 0 & 0 & 55&0 \\
%\hline
\end{tabular}
\vspace{0.6cm}
\caption{\label{table1} Properties of $S=-1$ resonances. $M_0$ and $\Gamma_0$ denote the pole mass and the width at the pole mass, respectively.}
\end{center}
\end{table}
\begin{table}[h]
\begin{center}
\begin{tabular}{|c||c|c|c|}
%\hline
channel & $a_0$[mb GeV$^2$] & $a_1$[GeV] & $a_2$ \\
\hline
$K^- p$ & 150 & 0.35 & 2 \\
\hline
$\bar{K}^0 n$ & 100 & 0.15 & 2\\
\hline
$\Lambda \pi^0$ & 130 & 0.25 & 3\\
\hline
$\Sigma^+ \pi^-$ & 600 & 0.1 & 2\\
\hline
$\Sigma^- \pi^+$ & 5000 & 0.1 & 3\\
\hline
$\Sigma^+ \pi^-$ & 2500 & 0.1 & 3\\
%\hline
\end{tabular}
\vspace{0.6cm}
\caption{\label{table2} Parameters for cross sections in $K^- p$ scattering
used for Eq.~(\ref{kmpex}).}
\end{center}
\end{table}

%\clearpage
%\newpage

%\clearpage
\newpage

\end{document}